# Quantum Puzzles in the Set Theoretic Metaworld of Heisenberg, Clauser, and Horne

Ian T. Durham

ABSTRACT

This paper follows up on a recent pre-print (Durham [2005]) by first deriving a set theoretic relationship between the generalized uncertainty relations and the Clauser-Horne inequalities. The physical, metaphysical, and metamathematical implications and problems are then explored. The discussion builds on previous work by Pitowsky [1994] and suggests that there is a fundamental problem in quantum correlation that could potentially lead to a paradox. It leaves open the question of whether the problem is in experiment, theory, or phenomena.

## 1 Introduction

In a recent pre-print I pointed out the numerous similarities between Bell-type inequalities and the generalized uncertainty relations, particularly in the context of conditional events (Durham [2005]). The thought experiment suggested in part 3 of that pre-print is a clear indication that this relationship is no coincidence. It turns out that a combination of set theory and logic can serve to show that the Bell-type inequalities known as the Clauser-Horne inequalities (or CH74 inequalities, Clauser & Horne [1974]) are a subset of the generalized uncertainty relations. Of course this begs the question of how realism is properly dealt with since it is an assumption at the heart of Bell's theorem. Interestingly enough, realism is not invoked at all in the process. However, in pondering the problem I have also made a distinctly philosophical suggestion regarding physical realism (to be distinguished from mathematical realism) and what its presence in quantum mechanics combined with my result might imply.

Much of the logic governing these various inequalities was actually first developed by George Boole in his analysis of 'conditions of possible experience.' Boole's priority was established a decade ago by Pitowsky [1994]. Pitowsky analyzed in detail the subsequent ways in which Boole's conditions could be violated and his analysis proves to be useful in supporting the CH74-Uncertainty relation developed in this preprint. In addition his algorithm for developing Boolean conditions is employed to produce a Boolean representation of the uncertainty relations. But first, let us derive the relationship that is the source of these philosophical problems. It is mathematical in the sense that it employs a bit of set theory, logic, and interval arithmetic. However, I make every attempt to define everything such that one need not be well-versed in any of the higher-order mathematics in order to understand the derivation. In addition I should note that I employ notation consistent with that employed by Gullberg [1998] simply in an effort to reach as wide an audience as possible.

## 2 Preliminaries

In their 1974 paper, Clauser and Horne offer a proof of the following: given six numbers, $x_1, x_2, y_1, y_2, X$, and $Y$, such that

$$0 \leq x_1 \leq X$$
$$0 \leq x_2 \leq X$$
$$0 \leq y_1 \leq Y$$
$$0 \leq y_2 \leq Y$$

then $-XY \leq f(x,y) \leq 0$ where

$$f(x,y) = x_1 y_1 - x_1 y_2 + x_2 y_1 + x_2 y_2 - Y x_2 - X y_1.$$

Once again, the last inequality is satisfied for *any* six numbers (the numbers do not have to represent *anything*) as long as the first set of four inequalities is satisfied. Set theoretically we can represent this most generally as

$$\chi = \{f(x,y) | -XY \leq f(x,y) \leq 0; 0 \leq x_n \leq X; 0 \leq y_n \leq Y\} \qquad (1).$$

For now I will refrain from defining $f(x,y)$ in these terms. Notice also that I have refrained from specifying whether the numbers are real or complex.
   Representing the uncertainty relations in an analogous way requires a bit more manipulation but is nonetheless possible. Let us begin by restating the generalized uncertainty relation as follows (see Liboff [1998]): given a commutation relation of the form $[A,B] = C$, then

$$\Delta A \cdot \Delta B \geq \tfrac{1}{2} |\langle C \rangle|.$$

The left side of the inequality represents the spread in the measurements of $A$ and $B$. Of course, a spread is an interval and each of these can be represented in set theoretic terms as $\Delta A = [a,a'] = \{x | a \leq x \leq a'\}$ and $\Delta B = [b,b'] = \{y | b \leq y \leq b'\}$. The product of the two intervals is another interval given by

$$\Delta A \cdot \Delta B = [a, a'] = [b, b'] = [\min(ab, ab', a'b, a'b'), \max(ab, ab', a'b, a'b')].$$

For simplicity's sake I will represent this combined interval as $\Psi = \{z | z_{\min} \leq z \leq z_{\max}\}$. To incorporate the right side of the generalized uncertainty relation let me first define $Z = \frac{1}{2}|\langle C \rangle|$. Therefore the generalized uncertainty relation becomes

$$\Psi = [z_{\min}, z_{\max}] \geq Z.$$

This says that the interval (or, more correctly, its length), $\Psi$, must be greater than or equal to half of $Z$. Working from the minimum requirement that the two are equal we can write $z_{\max} - z_{\min} = Z$ or $z_{\min} = z_{\max} - Z$. The interval can then be given as $z_{\max} - Z \leq z \leq z_{\max}$. With just a bit more algebra (I hope I haven't lost anyone yet) this can be rearranged (by subtracting out the $z_{\max}$ across the inequality) to read

$$-Z \leq z - z_{\max} \leq 0.$$

In doing this I have incorporated the minimum condition of the inequalty in the uncertainty relation into the interval. Basically, this encapsulates the minimum requirements of the uncertainty relation and incorporates both sides of the inequality. The interval can then be written in set theoretic terms as

$$\Psi = \{f(z) | -Z \leq f(z) \leq 0; z_{\min} \leq c \leq z_{\max}\} \qquad (2).$$

Our task, then, is to compare (1) and (2).

# 3 A Meta Challenge
## 3.1 Part I: Metamathematics

Let us call $\Psi$ the Uncertainty Set and $\chi$ the Clauser-Horne Set. Compare, once again, their similar structure:

| Uncertainty Set | Clauser-Horne Set |
|---|---|
| $\Psi = \{f(z) | -Z \leq f(z) \leq 0; z_{\min} \leq z \leq z_{\max}\}$ | $\chi = \{f(x,y) | -XY \leq f(x,y) \leq 0; x, y \in \mathbf{R}_+\}$ |

The first major point I'd like to make is that $\chi$ admits only positive reals while I have made no such stipulation for $\Psi$. This is because, as we well know, the uncertainty relation can and does admit complex numbers. In addition it is important to remember that $\Psi$ was derived for the *minimum* requirements set forth by the uncertainty relations. $\chi$ has no such stipulation. Note also that nowhere did I include the commutation relation as a stipulation for $\Psi$. This is an issue that needs to be addressed as well. In addition two metaphysical points that need to be addressed are the principle of local realism in relation to χ and causality in relation to Ψ (I do not purport to 'solve' local realism or causality, simply point out their relation to this problem).

I can actually deal with a few of these problems simultaneously. The commutation relation is usually formulated as a *pre*requisite to the uncertainty relations. However, there is nothing anywhere that requires *either* to be logically prior to the other. So, more correctly, the commutation relation is a *co*requisite of the uncertainty relations but only in the sense that if one exists so does the other. Therefore it is not necessary to include the commutation relation in the uncertainty set's list of conditionals. However, the commutation relation does tell us something very important about the uncertainty relations: that complex results are a possibility, i.e. $Z$ could be (and very often is) a complex number. So, though the commutation relation itself may not appear in the uncertainty set's list of conditionals it does lead to the condition that complex numbers are allowed. Of course, as I first pointed out, χ only allows positive reals. Once again, let us compare the two with the added conditions of complexity for Ψ.

| Uncertainty Set | Clauser-Horne Set |
|---|---|
| $\Psi = \{f(z) \mid -Z \leq f(z) \leq 0; z \in \mathbf{C}\}$ | $\chi = \{f(x,y) \mid -XY \leq f(x,y) \leq 0; x, y \in \mathbf{R}_+\}$ |

Note that I also eliminated the interval for $z$ just as I eliminated the intervals for $x$ and $y$ before. How can I justify this? Well, theoretically the intervals of $A$ and $B$ separately range from -∞ to +∞. They are only limited when they are taken together in relation to $Z$. So $z$ itself, which is just the interval product of $A$ and $B$, can range from -∞ to +∞ as well. The uncertainty relation only stipulates that whatever $z$ happens to be (and it can be anything), $f(z)$ must be between $-Z$ and 0.

Now, given the definitions I have just established for Ψ and χ, what sort of mathematical relationships can we draw between the two? First, since any real number can simply be represented as a complex number with the complex part set to zero, it is true that $\mathbf{R}_+ \subseteq \mathbf{C}$. Second, the real part of $Z$ can always be decomposed into a product of two smaller real numbers each of which could possibly be positive. We can say, then, that $|\text{Re}(Z)| \geq XY$. Therefore Ψ must have a greater cardinality than χ.

Before any definitive relational statement about Ψ and χ, we should be sure that the nature of the functions $f(x,y)$ and $f(z)$ don't prevent us from doing so. Ah, but that is the beauty of the representation. It is in fact *the values* of those functions that are defined on the given intervals in the defined sets *not the functions themselves*. And, since the *values* in χ are a subset of those in Ψ we really don't have problem. In addition, what is a function other than a mapping? Metamathematically one might say that functions are no more than a way to condense a series of numbers into a more convenient package. Two functions that produce identical numerical results must ultimately be identical. In addition, I should note that *the* Clauser-Horne inequalities (i.e. such that $-1 \leq f(x,y) \leq 0$) are merely an *element* (or subset) of χ just as the position-momentum or energy-time uncertainty relations are elements (or subsets) of Ψ. This distinction is important for the conclusion of the derivation which is that

$$\chi \subseteq \Psi.$$

Metamathematically, if one were to argue that it matters not what the numbers represent, simply that they allow for this behavior, we could stop there. The metamathematical argument here brings to mind the views of Wittgenstein and Hilbert. If the formalists and Sophists both in mathematics and physics were right then, by stripping everything to its barest essentials, the formalism represented in my manipulation of the given symbols under the stated rules, then I need not go any further. The problem, of course, resides with the late enigma Kurt Gödel. The incompleteness theorem, in particular the *second* incompleteness theorem, puts a serious crimp in the formalist program. In simplified terms, the second incompleteness theorem states "that the consistency of a formal system that contains arithmetic can't be formally proven within that system."[1] Whatever we might argue, it is enough to prod us into an examination of the physical (metamathematical?) and metaphysical problems I mentioned above. And, it turns out, such an examination will prove useful.

## 3.2 Part II: Metaphysics

Despite the results of Gödel (and also Turing)[2], the desire of Wittgenstein and Hilbert (not to mention Frege) is seductive, at least from a physical (as opposed to mathematical) standpoint. Eddington in his *Fundamental Theory* [1946] basically attempted to axiomatize (formalize) physics while simultaneously uniting relativity and quantum mechanics. Of course he wasn't generally successful, but his methods have left some useful nuggets (see Durham [2004]) and his mere attempt gives strict formalists some hope. Without taking a side in this argument (my own opinion might be surprising) I think it is worth attempting to attach a mathematical form to some of the metaphysical concepts represented in this problem.

Since $\chi$ is simply a generalized set theoretic representation of the Clauser-Horne inequalities, if we are to go beyond the strictly formal aspects of the system we must reconcile the result with local realism since it plays such an important role in any Bell-type inequality. A very generalized version of Bell's theorem essentially states that local realism implies some inequality. Experimental violation of Bell-type inequalities is taken to mean that local realism is incompatible with quantum mechanics. These inequalities are, in fact, just forms of Boole's 'conditions of possible experience.' As Pitowsky points out, "*none of Boole's conditions of possible experience can ever be violated when all the relative frequencies involved have been measured in a single sample.*"[3] Technically this is assuming that the numbers involve probabilities that are interpreted as relative frequencies (whether or not they actually are, they manifest as such in the case of repeatable, exchangeable, or independent events). However, the additional boundary conditions for $x_n$ and $y_n$ supplied by Clauser and Horne in the generalization release us from the requirement that the numbers must be interpreted as probabilities (falling between 0 and 1).  Thus, in the purely general form presented here χ does not necessarily need to be applied to the case of probabilities as relative frequencies.  Obviously from an experimental standpoint it might be difficult to manifest in any other way, but I am unaware of any conclusive proof of its impossibility.  In addition, Pitowsky points out that Boole's conditions can not only be derived from probability theory but also from propositional logic (Pitowsky [1994]) meaning probabilities are not *a priori* required even though that is the form in which Boole himself derived his conditions.  It is questionable, then, if it is the relative frequency interpretation that leads to the inequalities as Pitowsky points out.  If we use propositional logic instead (which certainly can be related to probabilities and relative frequencies but doesn't *have* to be) we still arrive at the same conditions with the same inequalities.

---

[1] Goldstein [2005], p. 183. See also the English translation of Gödel's theorem by Meltzer [1962].
[2] Turing's most seminal work in mathematical logic can be found in Davis [2004]. In addition most of Turing's writing (published and unpublished) can be found online at http://www.turing.org.uk.
[3] Pitowsky [1994], p.105.

Opening χ up to numbers other than probabilities still does nothing for local realism but returning to Pitowsky's point, Boole's conditions are not violated if all measurements are taken on a single sample.  In addition since Boole's conditions can be derived from very elementary assumptions the puzzling aspects of quantum mechanics are not only in the theory but also in the physical phenomena themselves (Pitowsky [1994]).  Physically the puzzling aspects are usually ascribed to interference.  As I will point out, despite its inherent metaphysical problems, when linked with a metamathematical interpretation the fog seems to clear a bit.  But how does the sampling affect the assumption of local realism and what does it have to say (if anything) about the puzzling nature of quantum mechanics?

It is time to phrase local realism in a more formal way.  Local realism, of course, is the combination of the Principle of Locality – that no information can travel faster than the speed of light – with the idea of realism – that measurements exist regardless of whether or not they've actually been measured, i.e. any correlation is *a priori* to the moment of measurement.  I take 'information' to mean anything – photons, massive particles, anything.  The requirement, then, that no information travel faster than the speed of light can be given the mathematical expression $\dot{I} \leq c$ where $\dot{I}$ is the rate of change of the 'information,' in a sense it's speed.  Is there a similar way to represent realism?  The question *may* be partially tied to the truth of realism.  For example, if realism is false – if these measurements are not *a priori* correlated – there shouldn't be any mathematical way to represent realism.  Of course, this conclusion is based on the assumption that all mathematics represents something physical which is an open question.  Actually, the circular nature of this problem needn't necessarily saddle us at the moment since the following suggestion makes no progress towards a solution – nor is it intended to.  It is simply a suggested formal way to represent physical realism in quantum mechanical situations.

The suggestion begins with the recognition that the puzzles in quantum mechanics that are discussed herein – namely violations of Boole's conditions of possible experience – are due to a phenomena most physicists (this one sometimes included, particularly in pedagogical situations) refer to as interference.  But as Pitowsky points out the very meaning of the term is at stake here (Pitowsky [1994]).  Whatever it is we might call it, it is mathematically represented in the formal equations of quantum mechanics by the imaginary parts of complex numbers.  From a measurement standpoint we can only measure the real parts of the complex numbers, nonetheless the effects of the imaginary portions of these numbers is clearly *qualitatively* observable.  In some sense one could argue they are also quantitatively observable, but the values themselves are not *directly* measurable. Let's take stock of just what this implies: we have numerical values that we can theoretically infer and whose effects can qualitatively be observed but that *cannot be directly measured*. If a way to directly measure these observables were to be developed it would seem to be evidence of physical realism since the *a priori* correlation between the real and imaginary parts of the complex values was established *prior* to any measurement of the imaginary portions having taken place. The argument falls well short of a proof but is a tantalizing suggestion that also provides us a 'temporary' way to formally represent realism: complex valued observables. Realism in this sense, then, is already present in $\Psi$ and $\chi$ (in the latter by the fact that $\mathbf{R}_+ \subseteq \mathbf{C}$ since real numbers are just complex numbers whose imaginary parts are 0).

Having put local realism on a formal (or semi-formal) footing, we are left with the metaphysical aspects of uncertainty. The problem that would prove the most difficult to reconcile with $\chi$ would be issues of causality. However, the uncertainty relations themselves, stripped bare of interpretation, are not inherently acausal. In fact Heisenberg originally stated that he had attempted to maintain causality and thereby preserve consistency with relativity (Jammer [1966]). His rationale was that uncertainty did not specifically restrict our knowledge of the *future*, rather that it restricted our knowledge of the *present*. In essence our knowledge of the present is incomplete and until found to be complete our knowledge of the future will be limited. As such, for the time being, if we assume uncertainty is causal it is correspondingly local.

Therefore we find that only one additional condition needs to be added to $\Psi$ and $\chi$, that of locality, since we have already accounted for realism through the complex number conditional $\mathbf{R}_+ \subseteq \mathbf{C}$. As such, we can write

| Uncertainty Set | Clauser-Horne Set |
|---|---|
| $\Psi = \{f(z) \mid -Z \leq f(z) \leq 0; z \in \mathbf{C}; \dot{I} \leq c\}$ | $\chi = \{f(x,y) \mid -XY \leq f(x,y) \leq 0; x,y \in \mathbf{R}_+; \dot{I} \leq c\}$ |

Given the formalized meta-conclusions, our statement that $\chi \subseteq \Psi$ now has a more solid foundation. I will discuss the implications of this in the next section.

## 4 Implications and Interpretations

From an interpretational standpoint, Von Neumann's idea that any observable that was uncertain had no value at all would seem to reduce everything – uncertainty included – to essentially binary results. Certainly this is an interpretation in the extreme. Nonetheless, as van Fraassen points out

> One recurring worry among philosophers is that the appearance of the term 'measurement' in the Born Rule bears its anthropocentric connotations essentially. That would mean that we cannot think of quantum theory as a putative autonomous description of the world in neutral physical terms and prospectively complete. In the jargon: if that were so, we could not be realists with respect to the theory, but only instrumentalists. This worry is much reinforced by 'philosophical' discussions by some of the great physicists who were involved in the development of quantum theory. [That question has been laid] to rest, since the requirements upon physical correlates of measurement involved no reference to us, to persons or consciousness, and not even to the macro-micro distinction.[4]

He goes on to make a very important point: that if the significance of the measurement itself is gone and we are left to the assignment of probabilities, "the *given* had better be handled very delicately." Fundamentally, this is reminiscent of precisely what Heisenberg had originally implied – uncertainty lies in our knowledge of the *present*. Van Fraassen's point is that it is not just our knowledge of the present that is the problem but *how we apply it*. For example, as Pitowsky pointed out, violations of inequalities representing Boole's conditions of possible experience are *logically impossible* if taken on a single sample. Perhaps a measurement on a single sample such as this is not possible, but perhaps it's not. Until conclusive evidence exists either way we should proceed with caution. In regard to the conclusion of my derivation, $\chi \subseteq \Psi$, this suggests that it would be best to ensure the firm mathematical foundation before proceeding to the metamathematical, physical, and metaphysical.

Just what would the conclusion that $\chi \subseteq \Psi$ mean for quantum mechanics? Let me first state emphatically that it would *not* mean that the CH74 inequalities were a subset or Boolean limit on the position-momentum or energy-time uncertainty relations. Clearly they deal with separate phenomena. However, the CH74 inequalities *are* elements of a set, $\chi$, that *is* a subset of the set $\Psi$, of which the uncertainty relations are elements. The uncertainty relations, however, are *not* elements of $\chi$. In addition, there is an uncertainty relation for Pauli spin matrices and also one for spin angular momentum, the same phenomena dealt with in the CH74 case. This suggests, then, that there ought to be an inequality similar in form to the CH74 inequalities that describes a binary (probably instrumental) case of simultaneous position-momentum or energy-time measurement. In fact Pitowsky provides the algorithm (see the Appendices of Pitowsky [1994]) by which this can be done. In my recent pre-print (Durham [2005]) I provided, in section 3, a thought experiment to match. The resulting inequality is trivial – $\{0 \leq p_1 + p_2 \leq 1\}$ – but nonetheless matches the thought experiment (there is no $p_{12}$ term for simultaneous *exact* measurements).

The metaphysical implications of this are that the underlying theorem of the CH74 inequalities, which is essentially Bell's theorem, appears to be an element of a set of theories that also includes Heisenberg's Uncertainty Principle as an element. This further suggests that there ought to be some theory that unites these two. In addition, any assumptions found (i.e. proven) to be fundamental to Bell's theorem must also be fundamental to the larger set that includes uncertainty. This would include realism, locality, causality, etc. Despite suggestions to the contrary (including my own regarding complex numbers in the previous section) realism is unlikely to be proven one way or another, partly due to incompleteness. Locality, on the other hand, is concretely measurable. In addition, a general conclusion of this work is that if violations of the CH74 and other Bell-type inequalities exist there ought to also exist violations of Heisenberg-type (uncertainty) inequalities, particularly if they are consistent with Boole's conditions of possible experience. However, in keeping with Boole's conditions violations could only be a logical possibility if taken on multiple samples since Pitowsky has shown violations on single samples to be logically *impossible*.

---
[4] van Fraassen [1991], p. 284.

Finally, if local realism is a theoretical necessity for the Clauser-Horne inequalities, my result suggests that it ought to be a theoretical necessity for the generalized uncertainty relations as well. But, before this suggestion elicits too many "groans," recall that the violations of Bell-type inequalities that purported to show the irreconcilability of local realism with quantum mechanics had to have been performed on the equivalent of multiple samples. Therefore, as long as the single sample requirement is met, there should be no inequality violation, and thus no contradiction between local realism and quantum mechanics.

## 5 Conclusions and Further Suggestions

The major conclusion of this work also happens to be a suggestion. As my discussion in the previous section has pointed out, building on the work of Pitowsky, my set theoretic results suggests a way to bring uncertainty into the fold of Boole's conditions of probable experience. It underscores the importance of Pitowsky's results by suggesting that the problem lies in the *experiment* rather than in the theory. Now, whether that problem can be overcome is another matter entirely. It is possible that measurements on single samples can actually be achieved in which case Boole's conditions can be thoroughly tested. Obviously any violation then would need to be meticulously analyzed to determine whether the problem lay in the experiment or in Boole's and Pitowsky's (and subsequently my) reasoning. But Pitowsky's work appears fairly solid and I personally doubt a violation following *strict* single sample guidelines could occur. With that said, if such a violation were to occur it might suggest a *fundamental limitation* on our very ability to make such a measurement. Perhaps it is just not possible to make such a measure on a single sample (though it might be possible to delude ourselves into thinking we succeeded).

Would this last point, then, be an instrumentation limit or something deep in the actual phenomenon? The latter point seems as if it would be paradoxical. Take the example of a pair of entangled electrons. If we were to succeed in sampling a singlet state to the point where we were absolutely confident in our experimental apparatus (i.e. *experimentally* certain that we had a *single* sample), and yet we obtain a violation nonetheless, it would suggest that the two electrons were *not* actually correlated in any way. But if we know from our initial conditions that the two electrons are entangled, they *have* to be correlated and therein lies the paradox. If one were to take a more formalist point of view and assume that Boole's conditions (and related derivations and inequalities) are all there is to the nature of the problem then one might be inclined to say that there simply cannot be a paradox and therefore we should try again. In essence a true formalist might say that there *never will be* a paradoxical situation like this. On the other hand a Platonist might be more skeptical, perhaps even going so far as to *expect* a paradox. But paradox's usually arise from an ill-formed argument and if one arose in this case both the experiment as well as the theory should be examined.

In any case, there seems to be a troubling problem with our usual conceptions of quantum correlation (which was problematic enough to begin with). It is distinctly possible that the resolution to this problem will evade us forever. My conclusion, however, should help clarify different aspects of the problem by providing additional conditions to be included in experiment.

*Department of Physics*
*Saint Anselm College*
*Manchester, New Hampshire*
*USA*
*idurham@anselm.edu*